\documentclass[pre,twocolumn,aps,floatfix,longbibliography,natbib]{revtex4-1}
\usepackage{pslatex,graphicx,dcolumn,bm,natbib,amssymb,amsmath,color}

\newcommand{\be}{\begin{equation}}
\newcommand{\ee}{\end{equation}}
\newcommand{\ket}[1]{|#1\rangle}
\newcommand{\braket}[2]{\langle #1|#2\rangle}

\begin{document}

\title{Motility-driven glass and jamming transitions in biological tissues}

\author
{
	Dapeng Bi$^{1,2}$, 
	Xingbo Yang$^{1,3}$,
	M. Cristina Marchetti$^{1,4}$,
	M. Lisa Manning$^{1,4}$ \\
		$^1$Department of Physics, Syracuse University, Syracuse, NY, USA \\
		$^2$Present address: Center for Studies in Physics and Biology, Rockefeller University, NY ,USA \\
		$^3$Present address: Department of Physics, Northwestern University, Evanston, IL \\
		$^4$Syracuse Biomaterials Institute, Syracuse, NY, USA
}

\begin{abstract}
Cell motion inside dense tissues governs many biological processes, including embryonic development and cancer metastasis, and recent experiments suggest that these tissues exhibit collective glassy behavior. To make quantitative predictions about glass transitions in tissues, we study a self-propelled Voronoi (SPV) model that simultaneously captures polarized cell motility and multi-body cell-cell interactions in a confluent tissue, where there are no gaps between cells. We demonstrate that the model exhibits a jamming transition from a  solid-like state to a fluid-like  state that is controlled by three parameters: the single-cell motile speed, the persistence time of single-cell tracks, and a target shape index that characterizes the competition between cell-cell adhesion and cortical tension. In contrast to traditional particulate glasses, we are able to identify an experimentally accessible structural order parameter that specifies the entire jamming surface as a function of model parameters. We demonstrate that a continuum Soft Glassy Rheology model precisely captures this transition in the limit of small persistence times, and explain how it fails in the limit of large persistence times. These results provide a framework for understanding the collective solid-to-liquid transitions that have been observed in embryonic development and cancer progression, which may be associated with Epithelial-to-Mesenchymal transition in these tissues.
\end{abstract}
\maketitle

Recent experiments have revealed that cells in dense biological tissues 
exhibit many of the signatures of glassy  materials, including caging, dynamical heterogeneities and viscoelastic behavior~\cite{Schoetz2013,Schoetz2008,Angelini_PRL_2010,Angelini_PNAS_2011,KaesNJP}. These dense tissues, where cells are touching one another with minimal spaces in between, are found in diverse biological processes including wound healing, embryonic development, and cancer metastasis. 

In many of these processes, tissues undergo an Epithelial-to-Mesenchymal Transition (EMT), where cells in a solid-like, well-ordered epithelial layer transition to a mesenchymal, migratory phenotype with less well-ordered cell-cell interactions~\cite{Theiry_EMT_review,Thompson_Newgreen_EMT_review}, or an inverse process, the Mesenchymal-to-Epithelial Transition (MET). Over many decades, detailed cell biology research has uncovered many of the signaling pathways involved in these transitions~\cite{MET_review,Nakaya_MET}, which are important in developing treatments for cancer and congenital disease. 

Most previous work on EMT/MET has focused, however, on properties and expression levels in single cells or pairs of cells, leaving open the interesting question of whether there is a collective aspect to these transitions: Are some features of EMT/MET generated by large numbers of interacting cells? Although there is no definitive answer to this question, several recent works have suggested that EMT might coincide with a collective solid-to-liquid jamming transition in biological tissues~\cite{Park_2015,Sadati-Fredberg-review, KaesNJP,Friedl_2014}.  Therefore, our goal is to develop a framework for jamming and glass transitions in a minimal model that accounts for both cell shapes and  cell motility, in order to make predictions that can quantitatively test this conjecture.

Jamming occurs in non-biological particulate systems (such as granular materials, polymers, colloidal suspensions, and foams) when their packing density is increased above some critical threshold,
and glass transitions occur when the fluid is cooled below a critical temperature. Over the past 20 years these phenomena have been unified by ``jamming phase diagrams"~\cite{LiuNagelReview,Trappe_nature_2001}. 

Building on these successes, researchers have recently used self-propelled particle (SPP) models to describe dense biological tissues~\cite{Szabo_PRE_2006, Belmonte_PRL_2008_cell_sorting,Henkes2011,Sepulvieda_et_al,Garcia_Gov_2015,Gupta_etal_2015}. These models are similar to those for inert particulate matter -- cells are represented as disks or spheres that interact with an isotropic soft repulsive potential -- but unlike Brownian particles in a thermal bath, self-propelled particles exhibit persistent random walks. 

SPP models typically exhibit a glass transition from a diffusive fluid state to an arrested sub-diffusive solid  that is controlled by (1) the strength of self-propulsion~\cite{Henkes2011,Ran_2013,Garcia_Gov_2015} and (2) the packing density $\phi$
~\cite{Henkes2011,Fily_2012,Ran_2013,Berthier_PRL_2014,Fily_SM_2014}.
Just like in thermal systems, 
a jamming transition occurs at a critical packing density $\phi_G$, but this critical density is altered by the persistence time of the random walks~\cite{Henkes2011,Fily_2012,Ran_2013,Berthier_PRL_2014,Fily_SM_2014}.

\begin{figure*}
\begin{center}
\includegraphics[width=1.5\columnwidth]{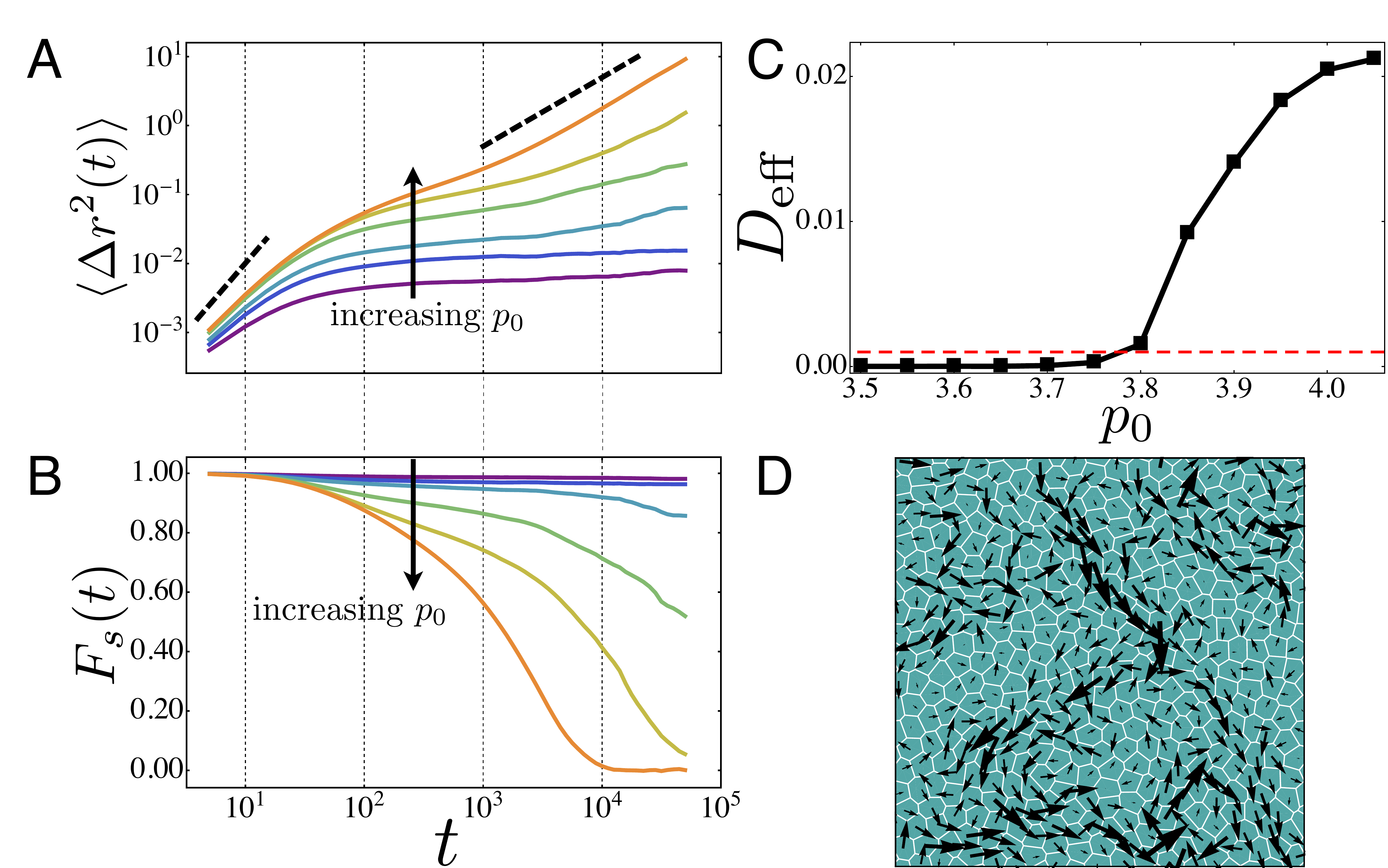}
\caption{
{\bf{Analysis of glassy behavior}}.
(A)
 The mean-squared displacement of cell centers for $D_r=1$ and $v_0=0.1$ and various values of $p_0$ (bottom to top: $p_0=3.5,3.65,3.7,3.75,3.8,3.85$) show the onset of dynamical arrest as $p_0$ is changed indicating a glass transition. The dashed lines indicate a slope of $2$(ballistic) and $1$(diffusive) on log-log plot. 
(B)
The self-intermediate scattering function at the same values of $p_0$ shown in (A) shows the emergence of caging behavior at the glass transition. 
(C) The effective self-diffusivity as function of $p_0$ at $v_0=0.1$. At the glass transition $D_{eff}$ becomes nonzero. 
(D)The cell displacement map in SPV model for a fluid state very close to the glass transition ($p_0=3.8$, $v_0=0.1$ and  $D_r=1$) over a time window $t = 10^4$  corresponding to the structural relaxation at which $F_s(t) \approx 1/2$.
}
\label{fig:glassy}
\end{center}
\end{figure*}

During many biological processes, however, a tissue remains at confluence (packing fraction equal to unity) while it changes from a liquid-like to a solid-like state or vice-versa. For example, in would healing, cells collectively organize to form a `moving sheet' without any change in their packing density~\cite{Kim_monoloayer_obstacle_2013}, and during vertebrate embryogenesis mesendoderm tissues are more fluid-like than ectoderm tissues, despite both having packing fraction equal to unity~\cite{Schoetz2013}. 

Recently, Bi and coworkers~\cite{bi_manning_nphys} have demonstrated that the well-studied vertex model for 2-D confluent tissues~\cite{ngai_honda,Farhadifar2007,Hufnagel2007,Staple2010,manning_2010,bi_softmatter} exhibits a rigidity transition in the limit of zero cell motility. Specifically, the rigidity of the tissue vanishes at a critical balance between cortical tension and cell-cell adhesion. An important insight is that this transition depends sensitively on cell shapes, which are well-defined in the vertex model.
While promising, vertex models are difficult to compare to some aspects of experiments because they do not incorporate cell motility.

In this work, we bridge the gap between the confluent tissue mechanics and cell motility by studying a hybrid between the vertex model and the SPP model, that we name Self-Propelled-Voronoi (SPV) model. A similar model was introduced by Li and Sun~\cite{Li_Sun_Biophys}, and cellular Potts models also bridge this gap~\cite{Szabo_Potts_2010, Kabla_CPM}, although glass transitions have not been carefully studied in any of these hybrid systems.

\section{The SPV model}
While the vertex model describes a confluent tissue as a polygonal tiling of space where the degrees of freedom are the vertices of the polygons, the SPV model identifies each cell only using the center ($\bm r_i$) of Voronoi cells in a Voronoi tessellation  of space (Dirichlet domains)~\cite{Dirichlet_1850}. 
The observation that Voronoi tessellations can describe cellular patterns in epithelial tissues was first proposed by Honda~\cite{honda_voronoi}.
For a tissue containing $N$-cells, the inter-cellular interactions are captured by a total energy which is the same as that in the vertex model.  Since the tessellation is completely determined by the $\{\bm r_i\}$, the  total tissue mechanical energy can be fully expressed as $E = E(\{\bm r_i\})$: 
\be
E= \sum_{i=1}^N \left[ K_A (A(r_i)-A_0)^2+ K_P (P(r_i)-P_0)^2 \right].
\label{eq:total_energy}
\ee
The term quadratic in cell area $A(r_i)$ results from a combination of cell volume incompressibility and the monolayer's resistance to height fluctuations~\cite{Hufnagel2007}.  The term involving the cell perimeter $P(r_i)$ originates from active contractility of the acto-myosin sub-cellular cortex (quadratic in perimeter) and effective cell membrane tension due to cell-cell adhesion and cortical tension (both linear in perimeter). This gives rise to an effective target shape index that is dimensionless: $p_0=P_0/\sqrt{A_0}$. $K_A$ and $K_P$ are the area and perimeter moduli, respectively.  For the remainder of this manuscript we assume $p_0$ is homogenous across a tissue, although heterogeneous properties are also interesting to consider~\cite{Wetzel_draft_2015}.

In the vertex model~\cite{bi_manning_nphys}, a rigidity transition takes place at a critical value of $p_0=p_0^* \approx 3.81$.
When $p_0<p_0^*$, cortical tension dominates over cell-cell adhesion and the energy barriers for local cell rearrangement and motion are finite.   The tissue then behaves as a elastic solid with finite shear modulus. When $p_0>p_0^*$, cell-cell adhesion dominates  and the energy barriers for local rearrangements vanish, resulting in zero rigidity and fluid-like behavior. 
While the energy functional for cell-cell interactions is identical in the vertex and SPV models, the two are truly distinct: the local minimum energy states of the vertex model are not guaranteed to be similar to a Voronoi tessellation of cell centers, although we do find them to be very similar in practice. Therefore, we are also interested in whether a rigidity transition in the SPV model coincides with the rigidity transition of the vertex model.

We define the effective mechanical interaction force experienced by cell $i$ as $\bm F_i=-\bm \nabla_i E$ (see Appendix~\ref{spv_forces} for details). In contrast to particle-based models, $\bm F_i $ is non-local and non-additive: $\bm F_i $ cannot be expressed as a sum of pairwise force between cells $i$ and its neighboring cells. Nevertheless, one can show that momentum is still precisely conserved by this energy functional in the absence of   the additional  self-propulsion forces introduced below.

In addition to  $\bm F_i$, cells can also move due to self-propelled motility. Just as in SPP models, we assign a polarity vector $\hat{\bm n}_i=(\cos\theta_i,\sin\theta_i)$ to each cell; along $\hat{\bm n}_i$  the cell exerts a self-propulsion force with constant magnitude $v_0/\mu$, where $\mu$ is the mobility (the inverse of a frictional drag).  Together these forces control the over-damped equation of motion of the cell centers $\bm r_i$
\be
\frac{d \bm r_i}{dt} =\mu \bm F_i +  v_0 \hat{\bm n}_i.
\label{equation_of_motion}
\ee

\begin{figure*}
\begin{center}
\includegraphics[width=1.5\columnwidth]{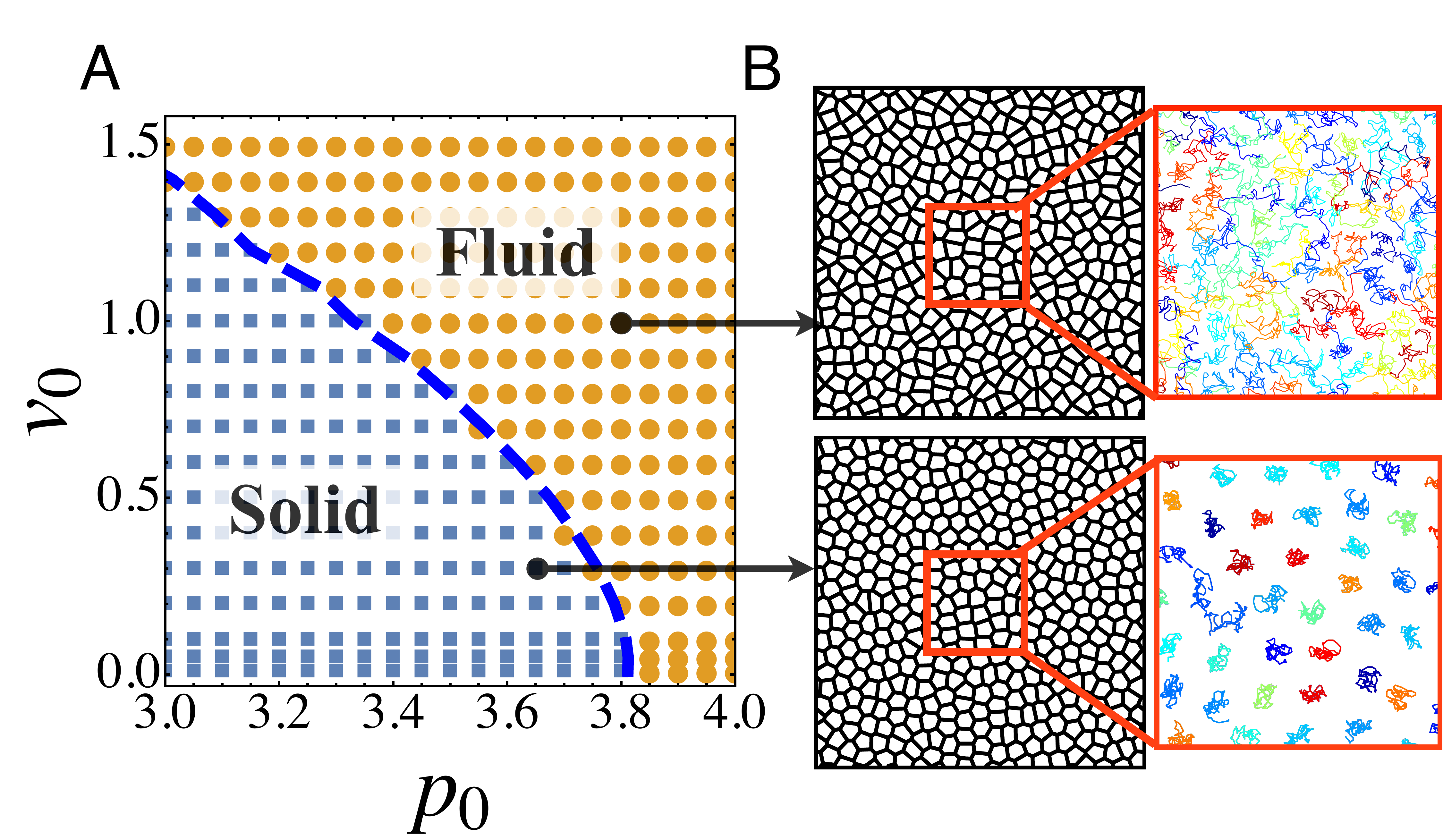}
\caption{
(A) Glassy phase diagram for confluent tissues as function of cell motility $v_0$ and target shape index $p_0$ at fixed $D_r=1$. 
Blue data points correspond to solid-like tissue with vanishing $D_{eff}$; orange points correspond to flowing tissues (finite $D_{eff}$). 
The dynamical glass transition boundary also coincides with the locations in phase space where the structural order parameter $q = \left\langle p/\sqrt{a} \right\rangle = 3.81$ (dashed line). In the solid phase, $q \approx 3.81$ and $q>3.81$ in the fluid phase. 
(B) Instantaneous tissue snapshots  show the difference in cell shape across the transition. 
Cell tracks also show dynamical arrest due to caging in the solid phase and diffusion in the fluid phase. 
}
\label{fig:phase_diagram}
\end{center}
\end{figure*}

The polarity is a minimal representation of the front/rear characterization of a motile cell~\cite{Szabo_Potts_2010}. While the precise mechanism for polarization in cell motility is an area of intense study, here we model its dynamics as a unit vector that undergoes random rotational diffusion
\be
\begin{split}
\partial_t \theta_i &= \eta_i(t) \\
\langle \eta_i(t)\eta_j(t') \rangle &= 2D_r\delta(t-t')\delta_{ij} 
\label{polarization_equation_of_motion}
\end{split}
\ee
where $\theta_i$ is the polarity angle that defines $\hat{\bm n}_i$, and $\eta_i(t)$ is a white noise process with zero mean and variance $2 D_r$. 
 The value of angular noise $D_r$ determines the memory of stochastic noise in the system, giving rise to a persistence time scale $\tau=1/D_r$ for the polarization vector $\mathbf{\hat{n}}$. For small $D_r \ll 1$, the dynamics of $\mathbf{\hat{n}}$ is more persistent than the dynamics of the cell position. At large values of $D_r$, i.e. when $1/D_r$ becomes the shortest timescale in the model,  Eq.~\eqref{equation_of_motion} approaches simple Brownian motion.

The model can be non-dimensionalized by expressing all lengths in units of $\sqrt{A_0}$ and time in units of $1/(\mu K_A A_0)$.   There are three remaining independent model parameters: the self propulsion speed $v_0$, the cell shape index $p_0$, and the rotational noise strength $D_r$.  
We simulate a confluent tissue under periodic boundary conditions with constant number of $N=400$ cells (no cell divisions or apoptosis) and assume that the average cell area coincides with the preferred cell area, i.e. $\langle A_i \rangle = A_0$. This approximates a large confluent tissue in the absence of strong confinement.  We numerically simulate the model using molecular dynamics by performing $10^5$ integration steps at step size $\Delta t=10^{-1}$ using Euler's method.
A detailed description of the SPV implementation can be found in the Appendix Sec.~\ref{spv_forces}.

\section{Characterizing glassy behavior}

We first characterize the dynamics of cell motion within the tissue by analyzing the mean-squared displacement (MSD) of cell trajectories. In Fig.~\ref{fig:glassy}(a), we plot the MSD as function of time, for tissues at various values of $p_0$ and fixed $v_0=0.1$ and $D_r = 1$. The MSD exhibits ballistic motion ( slope close to $2$ on a log-log plot) at short times, and plateaus at intermediate timescales. The plateau is an indication that cells are becoming caged by their neighbors. For large values of $p_0$, the MSD eventually becomes diffusive (slope =1), but as $p_0$ is decreased, the plateau persists for increasingly longer times. This indicates dynamical arrest due to caging effects and broken ergodicity, which is a characteristic signature of glassy dynamics.  

Another standard method for quantifying glassy dynamics is the self-intermediate scattering function~\cite{van_Hove_1954}:
$
F_s(k,t) = \left\langle e^{i \vec{k}\cdot \Delta \vec{r}(t)} \right\rangle .
$
Glassy systems possess a broad range of relaxation timescales, which show up as a long plateau in $F_s(t)$ when it is analyzed at a lengthscale $q$ comparable to the nearest neighbor distance.  Fig~\ref{fig:glassy} (b) illustrates precisely this behavior in the SPV model, when
 $| \vec{k} |=\pi/r_0$, where $r_0$ is the position of the first peak in the pair correlation function. The average $\langle ... \rangle$ is taken temporally as well as over angles of $\vec{k}$. $F_s(t)$ also clearly indicates that there is a glass transition as a function of $p_0$: at high $p_0$ values $F_s$ approaches zero at long times, indicating that the structure is changing and the tissue behaves as a viscoelastic liquid. At lower values of $p_0$, $F_s$ remains large at all timescales, indicating that the structure is arrested and the tissue is a glassy solid.
Fig~\ref{fig:glassy} (d) demonstrates that at the structural relaxation time, the cell displacements show collective behavior  across large lengthscales suggesting strong dynamical heterogeneity. This is strongly reminiscent of the `swirl' like collective motion seen in experiment in epithelial monolayers~\cite{Angelini_PRL_2010,Angelini_PNAS_2011, Poujade_2007,Silberzan2010, Garcia_Gov_2015}.

\subsection{A dynamical order parameter for the glass transition}
Although the phase space for this model is three dimensional, we now study the model at a fixed value of $D_r=1$.

We then search for a dynamical order parameter that distinguishes between the glassy and fluid states as a function of the two remaining model parameters,$(v_0,p_0)$. A candidate order parameter is the self-diffusivity $D_s$: 
$
D_s = \lim_{t \to \infty}  \langle {\Delta r(t)}^2 \rangle / (4t).
$
For practicality, we calculate $D_s$ using simulation runs of $10^5$ time steps, chosen to be much longer than the typical caging timescale in the fluid regime. 
We present the self-diffusivity in units of $D_0=v_0^2/(2D_r)$, which is the free diffusion constant of an isolated cell.  $D_{eff}=D_s/D_0$ then serves as an accurate dynamical order parameter that distinguishes a fluid state from a solid (glassy) state in the space of $(v_0,p_0)$, matching the regimes identified using the MSD and $F_q$. In Fig.~\ref{fig:phase_diagram}, the fluid region is characterized by a finite value of $D_{eff}$ and $D_{eff}$ drops below a noise  floor of $\sim 10^{-3}$ as the glass transition is approached. In practice, we label materials with $D_{eff}>  10^{-3}$ as fluids indicated by an orange dot, and those with $D_{eff}\le  10^{-3}$ as solids indicated by blue squares. Importantly, we find that the SPV model in the limit of zero cell motility shares a rigidity transition with the vertex model~\cite{bi_manning_nphys} at $p_0 \approx 3.81$, and that this rigidity transition controls a line of glass transitions at finite cell motilities. Typical cell tracks (Fig.~\ref{fig:phase_diagram}) clearly show caging behavior in the glassy solid phase.

\subsection{Cell shape is a structural order parameter for the glass transition}
In glassy systems it can be difficult to experimentally  distinguish between a truly dynamically arrested state and a state with relaxation times longer than the experimental time window. Similarly,  in tissues it is experimentally challenging to quantify a glass transition through the measurement of a dynamical quantity such as the diffusivity  $D_s$. Identifying a static quantity that directly probes the mechanical properties of a tissue  would therefore be a powerful tool for experiments. Puliafito \emph{et al.} have suggested that shape changes accompany dynamical arrest in proliferating tissues~\cite{Puliafito_2012}.  Similarly, a structural signature based on cell shapes -- 
the shape index $q = \left\langle p/\sqrt{a} \right\rangle$ -- was previously shown to be an excellent order parameter for the confluent tissue rigidity transition in the vertex model~\cite{Park_2015}. In a model where cells were not motile ($v_0 = 0$) we found that when $p_0 < 3.813$, $q$ is constant $\sim 3.81$ and when $p_0 > 3.81$ $q$ grows linearly with $p_0$. Quite surprisingly, we found that the prediction of $q=3.813$ works perfectly in identifying a jamming transition in \emph{in-vitro} experiments involving primary human tissues, where cells are clearly motile ($v_0 \neq 0$)~\cite{Park_2015}. At that time, we did not understand why the $v_0=0$ theory worked so well for these tissues.

The prediction of a solid-liquid transition in the SPV model presented here provides an explanation for this observation.We find that $q$ (which can be easily calculated in experiments or simulations from a snapshot) can be used as a structural order parameter for the glass transition for all values of $v_0$, not just at $v_0=0$. Specifically, the boundary defined by $q=3.813$, shown by the  blue dashed line in~Fig.~\ref{fig:phase_diagram}(A) coincides extremely well with the glass transition line obtained using the dynamical order parameter, shown by the round and square data points. The insets to Fig.~\ref{fig:phase_diagram} also illustrate typical cell shapes: cells are isotropic on average in the solid phase and anisotropic in the fluid phase.   This highlights the fact that $q$ can be used as a structural order parameter for the glass transition at all cell motilities, providing a powerful new tool for analyzing tissue mechanics. 

\section{A three-dimensional jamming phase diagram for tissues}
Having studied the glass transition as function of $v_0$ and $p_0$ at a large value of $D_r$, we next investigate the full three-dimensional phase diagram by characterizing the effect of $D_r$ on tissue mechanics and structure.  $D_r$ controls the persistence time $\tau = 1/D_r$ and persistence length or P\'{e}clet number $Pe \sim v_0/D_r$ of cell trajectories; smaller values of $D_r$ correspond to more persistent motion.

\begin{figure*}
\begin{center}
\includegraphics[width=2\columnwidth]{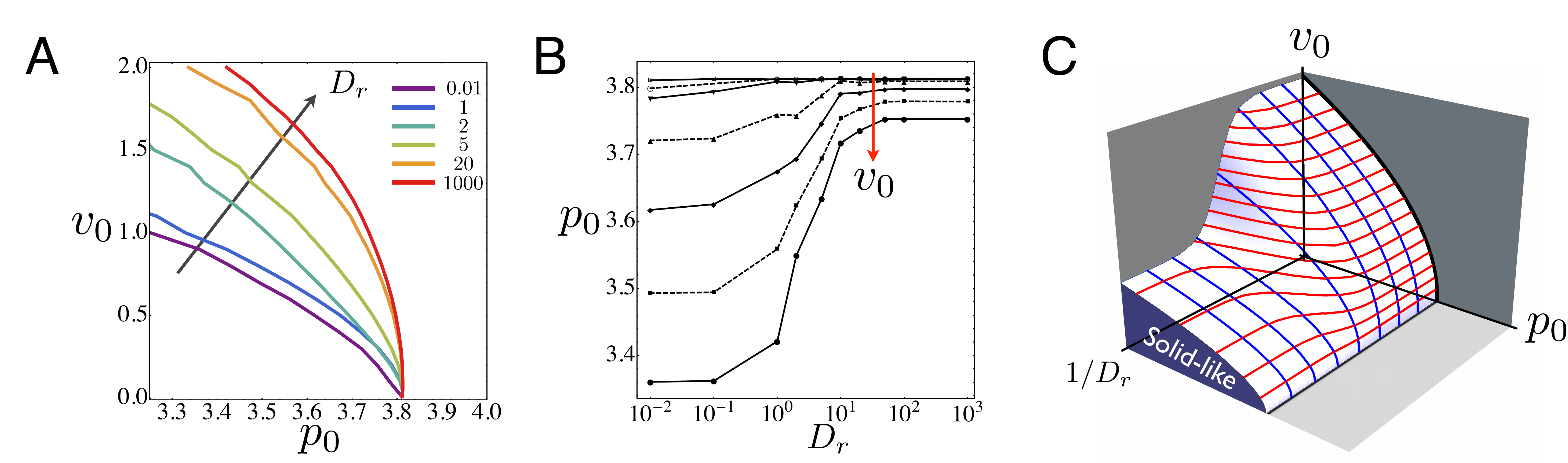}
\caption{
\textbf{(A)} 
The glass transition in $v_0-p_0$ phase space shifts as the persistence time changes. Lines represent the glass transition identified by the structural order parameter $q= 3.81$. The phase boundary collapse to a single point at $p_0^*=3.81$, regardless of $D_r$, in the limit $v_0 \rightarrow 0$.
\textbf{(B)} 
The glass transition in $p_0-D_r$ phase space shifts as a function of $v_0$ (from top to bottom: $v_0=0.02,0.08,0.14,0.2,0.26$) For large $v_0$ there is a crossover in the behavior at $D_r \sim \mu K_A A_0 = 1$, as discussed in the main text.
\textbf{(C)}
The phase boundary between solid and fluid as function of motility $v_0$, persistence $1/D_r$ and $p_0$ which is tuned by cell-cell adhesion can be organized into a schematic 3D phase diagram. 
Red lines on the surface correspond to iso-$v_0$ contours and blue lines correspond to iso-$D_r$ contours. 
}
\label{fig:dr}
\end{center}
\end{figure*}

In Fig.~\ref{fig:dr}(A), we show several 2D slices of the three dimensional jamming boundary. Solid lines illustrate the phase transition line identified by the structural order parameter $q = 3.813$ as function of $v_0$ and $p_0$ for a large range of $D_r$ values (from $10^{-2}$ to $10^3$). (In {\it Appendix~\ref{Dr_dep_boundary} } we demonstrate that the structural transition line $q = 3.813$ matches the dynamical transition line for all studied values of $D_r$.) In contrast to results for particulate matter~\cite{Berthier_PRL_2014}, this figure illustrates that the glass transition lines meet at a single point ($p_0=3.81$) in the limit of vanishing cell motility, regardless of persistence.

Fig.~\ref{fig:dr}(B) shows an orthogonal set of slices of the jamming diagram, illustrating how the phase boundary shifts as function of $p_0$ and $D_r$ at various values of $v_0$. This highlights the interesting result that a solid-like material at high value of $D_r$ can be made to flow simply by lowering its value of $D_r$.  The crossover in behavior at large $v_0$ occurs when the persistence time $1/D_r$ is approximately equal to the viscous relaxation time $1/(\mu K_A A_0) = 1$. 

These slices can be combined to generate a three-dimensional jamming phase diagram for confluent biological tissues, shown in~Fig.~\ref{fig:dr}(C). This diagram provides a concrete, quantifiable prediction for how macroscopic tissue mechanics depends on single-cell properties such as motile force, persistence, and the interfacial tension generated by adhesion and cortical tension. 

We note that~Fig.~\ref{fig:dr}(C) is significantly different from the jamming phase diagram conjectured by Sadati et al~\cite{Sadati-Fredberg-review}, which was informed by results from adhesive particulate matter~\cite{Trappe_nature_2001}. For example, in particulate matter adhesion enhances solidification, while in confluent models adhesion increases cell perimeters/surface area and enhances fluidization. In addition, we identify ``persistence" as a new axis with a potentially significant impact on cell migration rates in dense tissues. 

\begin{figure}[ht]
\begin{center}
\includegraphics[width=1\columnwidth]{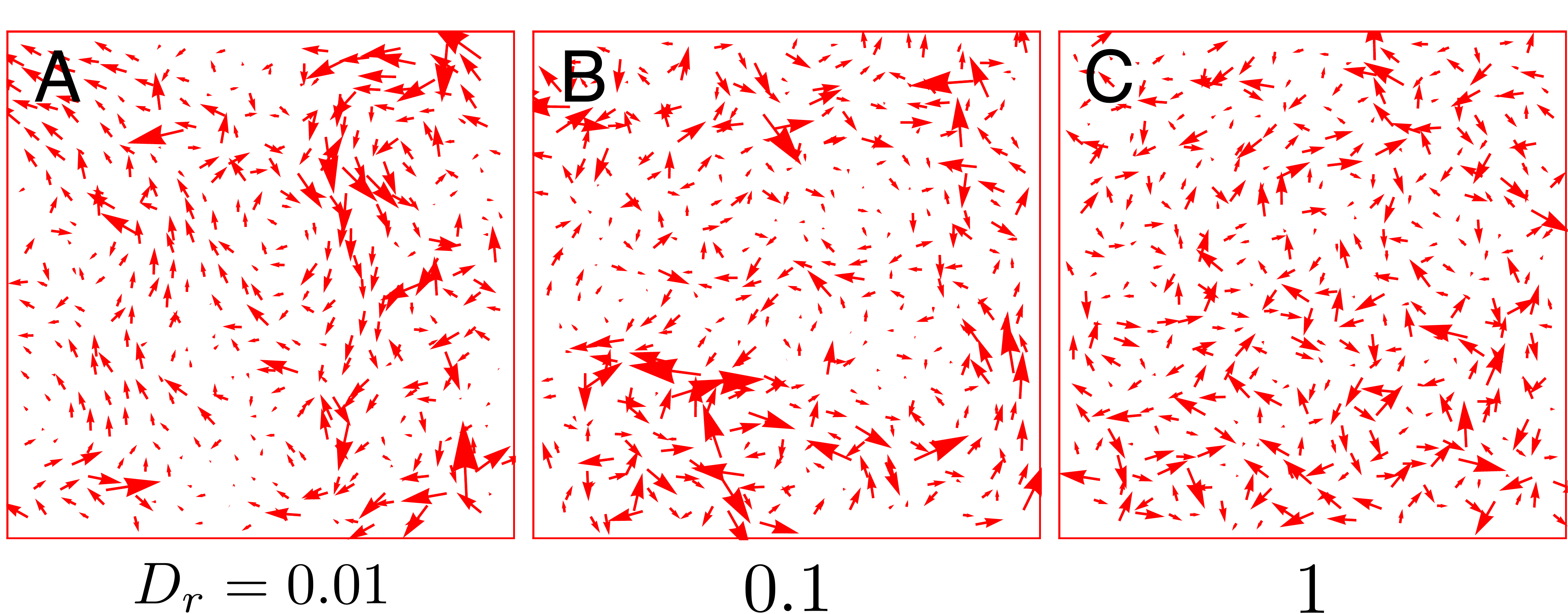}
\caption{
(A-C) Instantaneous cell displacements at $p_0=3.65$ and $v_0=0.5$. They are different from the displacements shown in Fig.~\ref{fig:glassy}(D) which are averaged over the structural relaxation timescale. (A) At the lowest value of $D_r = 0.01$, the cells are able to flow by collectively displacing along the `soft' modes of the system (Appendix.~\ref{soft_mode_analysis}). (B) At $D_r = 0.1$, collective displacements are less pronounced. (C) For $D_r = 1$ and larger, the displacements appear disordered and uncorrelated.
}
\label{fig:inst_disp}
\end{center}
\end{figure}

To better understand why persistence is so important in dense tissues, we first have to characterize the transitions between different cellular structures. In the limit of zero cell motility, the system can be described by a potential energy landscape where each allowable arrangement of cell neighbors corresponds to a metastable minimum in the landscape. There are many possible pathways out of each metastable state: some of correspond to localized cell rearrangements, while others correspond to large-scale collective modes.  The maximum energy required to transition out of a metastable state along each pathway is called an energy barrier~\cite{bi_softmatter}.

We observe that tissue fluidity can increase drastically with increasing $D_r$ at finite cell speeds.  This suggests that different pathways (with lower energy barriers) must become dynamically accessible at higher values of $D_r$.  

One hint about these pathways comes from the instantaneous cell displacements, shown for different values of $D_r$ in  Fig.~\ref{fig:inst_disp}.  At high values of $D_r$, ($p_0=3.78$, $v_0=0.1$) the instantaneous displacement field is essentially random and largely uncorrelated, as shown in Fig.~\ref{fig:inst_disp}, and the material is solid-like. There is no collective behavior among cells, and each cell `rattles' independently near its equilibrium position. 

However, as $D_r$ is lowered, the instantaneous displacement field becomes much more collective (Fig.~\ref{fig:inst_disp}) and the tissue begins to flow, presumably because these collective displacement fields correspond to pathways with lower transition energies.

Two obvious questions remain: How does a lower value of $D_r$ generate more collective instantaneous displacements? Why should collective instantaneous displacements generically have lower energy barriers?
The first question can be answered by extending ideas first proposed by Henkes, Fily and Marchetti~\cite{Henkes2011} to explain why motion in self-propelled particle models seems to follow the `soft modes' of a solid. 
This argument is based on a simple, yet powerful observation: in the limit of zero motility ($v_0=0$), a solid-like state will have a well-defined set of normal modes of vibration (with frequencies $\{\omega_\nu\}$), and a corresponding set of eigenvectors ($\{\hat{e}_{\nu}\}$) that forms a complete basis. 
At higher motilities ($v_0>0$) near the glass transition, the motion of particles in the system can be expanded in terms of the eigenvectors. As discussed in  Appendix~\ref{soft_mode_analysis}, one can use this observation to show that in the limit of $D_r\to 0$, motion along the lowest frequency eigenmodes is amplified -- the amplitude along each mode is proportional to $1/\omega_\nu^2$. These low-frequency normal modes are precisely the collective displacements observed for low $D_r$.  

The second question is more difficult to answer because it is impossible to enumerate all of the possible transition pathways and energy barriers in a disordered material.  However, a partial answer comes from recent work in disordered particulate matter showing that low-frequency normal modes do have significantly lower energy barriers~\cite{Xu_EPL_2010, Manning_Liu_PRL_2011} than higher frequency normal modes. A similar analysis could potentially be performed in vertex or SPV models.

\section{A continuum model for glass transitions in tissues}

Although continuum hydrodynamic equations of motion have been developed by coarse graining SPP models in the dilute limit, there is no existing continuum model for a dense active matter system near a glass transition. Here we propose that a simple trap~\cite{bouchaud_trap} or Soft Glassy Rheology (SGR)~\cite{Sollich_SGR_1998} model provides an excellent continuum approximation for the phase behavior in the large $D_r$ Brownian regime, but fails in the small $D_r$ limit.

For large $D_r$ it is known that particle behave like Brownian particles with an effective temperature $T_{eff}=v_0^2/2\mu D_r$~\cite{Fily_2012}. This mapping becomes exact when $D_r\rightarrow\infty$ at fixed ``effective inertia'' $(\mu D_r)^{-1}$~\cite{Fily_SM_2014}. In other words, like in granular systems~\cite{Abate_Durian_effective_temp, Cecconi_PRL_2003}, the effective temperature in SPP is dominated by kinetic effects. Guided by this result we conjecture that in our model the temperature also scale quadratically with the velocity,
\be
T_{eff} \propto c v_0^2.
\label{Teff}
\ee
Physically, this effective temperature gives the amount of energy available for individual cells to vibrate within their cage or `trap'.  

The next important question is how to characterize the `trap depths', or energy barriers between metastable states.  In the Brownian regime (large $D_r$) there is no dynamical mechanism for the cells to organize collectively, and therefore a reasonable assumption is that the rearrangements are small and localized. 


\begin{figure}[ht]
\begin{center}
\includegraphics[width=1\columnwidth]{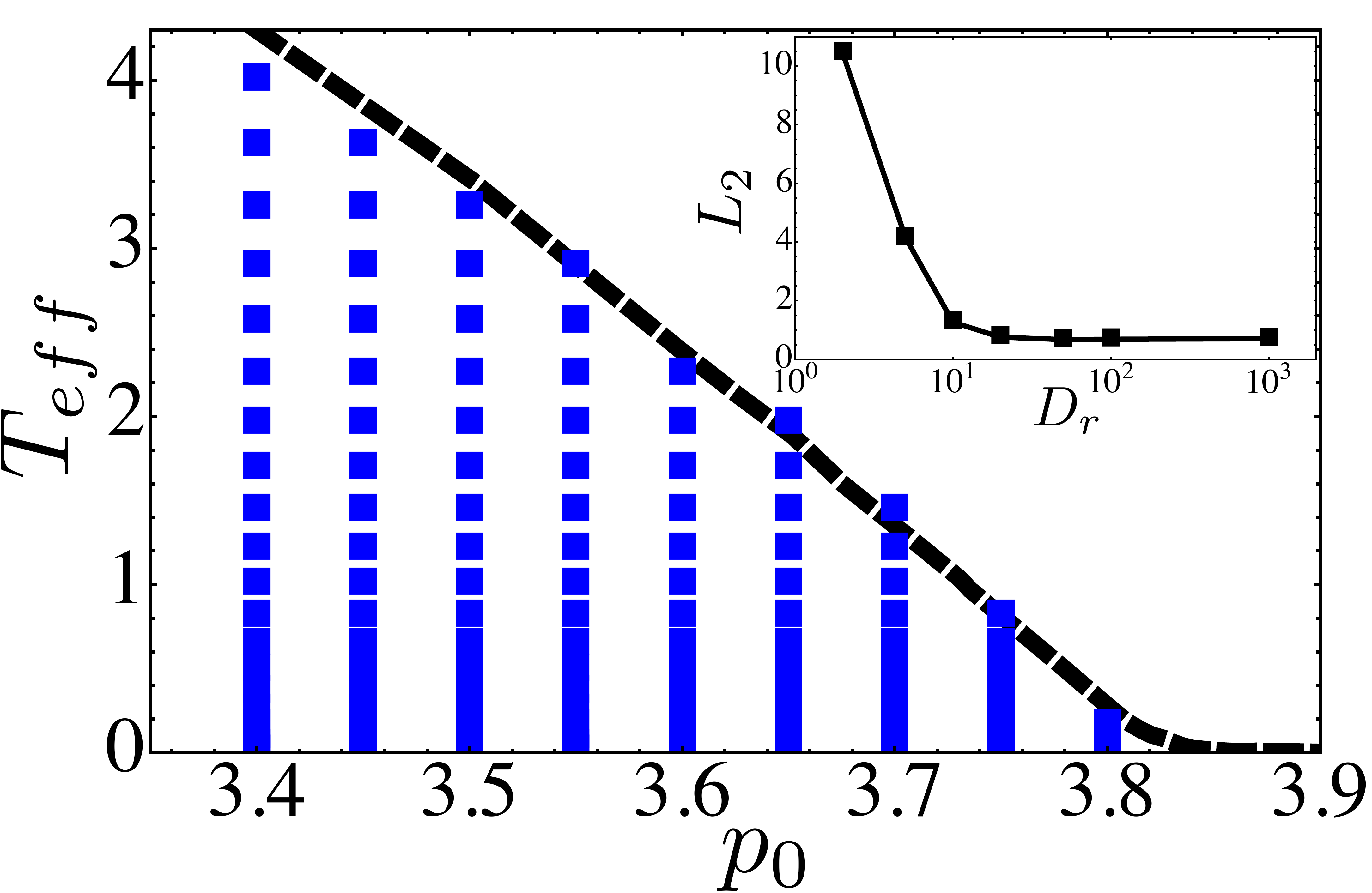}
\caption{
Comparison between SPV glass transition and an analytic prediction based on a Soft Glass Rheology (SGR) continuum model.  The dashed line corresponds to an SGR prediction with no fit parameter based on previously measured vertex model trap depths~\cite{bi_softmatter}.  Data points correspond to SPV simulations with $D_r = 10^{-3}$ and where we have defined $T_{eff} = c v_0^2$ with  $c = 0.1$ as the best-fit normalization parameter. Blue points correspond to simulations which are solid-like, with $D_{eff} < 10^{-3}$, and the boundary of these points define the observed SPV glass transition line. (Inset) $L_2$ difference between SPV glass transition line (at best-fit value of $c$) and the predicted SGR transition line at various values of $D_r$. The SGR prediction based on localized T1 trap depths works well in the high $D_r$ limit, but not in the low $D_r$ limit.
}
\label{fig:sgr}
\end{center}
\end{figure}

In~\cite{bi_softmatter}, some of us explicitly calculated the statistics of energy barriers for localized rearrangements in the equilibrium vertex model. In the 2D vertex model, one can show that localized rearrangements must occur via so-called T1 transitions~\cite{Weaire_book}. Using a trap model~\cite{bouchaud_trap} or Soft Glassy Rheology(SGR)~\cite{Sollich_SGR_1998} framework, we were able to use these statistics to generate an analytic prediction, with no fit parameters, for the glass transition temperature $T_g$ as function of $p_0$.

To see if the SGR prediction for the glass transition holds for the SPV model in the large $D_r$ limit, we simply overlay the data points corresponding to glassy states from the SPV model with the glass transition $T_g$ line predicted in~\cite{bi_softmatter}. There is one fitting parameter $c$ that characterizes the proportionality constant in Eq.~\ref{Teff}. Fig~\ref{fig:sgr}, shows that the SPV data for $D_r=10^3$ is in excellent agreement with our previous SGR prediction. 
Because $T_{eff}\sim v_0^2$, and the glass transition line scales as $T_g \sim p_0^* - p_0$ for  $p_0 \ll p_0^*$ and $D_r \to 0$, the glass transition line scales as $v_0 \sim {(p_0^*-p_0)}^{0.5}$ in those limits.

The reason the effective temperature SGR model works here is that, like in SPP models of spherical active Brownian colloids, the angular dynamics of each cell evolves independently of cell-cell interactions and of the angular dynamics of other cells.  An additional alignment interaction that couples the angular and translational dynamics may therefore modify this behavior.

To our knowledge, this is the first time that a SGR / trap model prediction has been precisely tested in any glassy system. This is because, unlike most glass models, we can enumerate all of the trap depths for localized transition paths in the vertex model. 

However, for small values of $D_r$, we have shown that cell displacements are dominated by collective normal modes, and therefore the energy barriers for localized T1 transitions are probably irrelevant in this regime. The inset to Fig~\ref{fig:sgr} shows the deviation ($L^2$-norm) between glass transition lines in the SPV model and T1-based SGR prediction as a function of $D_r$. We see that the SGR prediction fails in the small $D_r$ limit, as expected. A better understanding of the energy barriers associated with collective modes will be required to modify the theory at small $D_r$.

\section{Discussion and Conclusions}

We have shown that a minimal model for confluent tissues with  cell motility exhibits glassy dynamics at constant density.
 This model allows us to make a quantitative prediction for how the fluid-to-solid/jamming transition in biological tissues depends on parameters such as the cell motile speed, the persistence time associated with directed cell motion, and the mechanical properties of the cell (governed by adhesion and cortical tension). We define a simple, experimentally accessible structural order parameter -- the cell shape index -- that robustly identifies the jamming transition, and we show that a simple analytic model based on localized T1 rearrangements precisely predicts the jamming transition in the large $D_r$ limit. We also show that this prediction fails in the small $D_r$ limit, because the instantaneous particle displacements are dominated by collective normal modes.

This model makes several experimentally-verifiable predictions for cell shape and tissue mechanics:
\begin{itemize}

\item{{\em The order parameter $q=3.81$ is a structural signature for the glass transition, even in tissues with significant cell motility or dynamics.}  This prediction has already been tested in epithelial lung tissue~\cite{Park_2015}, but it should be much more broadly applicable. We have performed a rudimentary shape analysis of a small number of images from other systems that have been previously published, including proliferating MCDK monolayers~\cite{Deforet_ncomm_2014} and convergent extension in fruit fly development~\cite{matthias_elife} and found that the shapes are consistent with this prediction. A much more careful analysis with full data sets should be performed to further validate this prediction or understand where it breaks down.}

\item{ {\em  In the limit of vanishing cell motility, shape and pressure fluctuations should vanish when the jamming transition is approached from the solid side, and remain zero in the fluid.} A finite motility $v_0$ will induce such fluctuations in the fluid phase, as confirmed by preliminary  calculation of cellular stresses and pressure in the SPV model~\cite{SPV_stress_paper}. This could be studied by combining measurement of cell shape fluctuations with traction force  microscopy (TFM) in wound healing assays.  After locating the glass transition by imaging cell shape changes, it may be possible to extract information on cell motility $v_0$ from cellular stresses and pressure inferred from TFM in the fluid phase near the glass transition. This suggests that one may estimate cell motility by examining the changes in cellular stresses and pressure  in the cell monolayer near the unjamming transition and assuming that the local velocity of the monolayer is very small just above the transition. The latter assumption can also be verified independently via particle image velocimetry (PIV). }

\item{
{\em 
	Cell proliferation, so far neglected in our model, causes an increase in cell number density in confluent tissues. 
}
Often this is accompanied by a reduction in individual cell motility $v_0$, via contact inhibition of locomotion. In cases where this is the dominant effect and changes to the ratio between $A_0$ and $P_0$ are negligible, our work predicts
 that proliferation would drive the system towards jamming.
This is consistent with existing reports in the literature~\cite{Deforet_ncomm_2014}, although more work is required to test the prediction carefully. In tissues where $v_0$ remains low at all times~\cite{Puliafito_2012}, our model predicts that proliferation can either cause jamming or unjamming, depending on whether cell divisions are oriented in such a way to decrease or increase cell shape anisotropy.}

\item{{\em Spatial correlations and fluctuations of the cell displacement field, such as swirl€™ sizes~
\cite{Poujade_2007,Silberzan2010,Angelini_PRL_2010,Angelini_PNAS_2011,Garcia_Gov_2015}
, should grow as a tissue approaches the glass transition from the fluid side.} 
Very recently, a similar prediction for displacements and correlation lengths based on a particle-based model has been verified in one cell type~\cite{Garcia_Gov_2015}. The SPV model, which makes predictions for cell shapes in addition to displacements and correlation lengths,
could be tested simply by compiling detailed statistics about cell shapes and cell motion in epithelial monolayers.
}

\end{itemize}

  Although all of the work presented here focuses on the SPV model, which tracks cell centers and therefore has only two degrees of freedom per cell, we found that in the limit of zero cell motility it exhibits the same rigidity transition as the vertex model which has two degrees of freedom per vertex. We have also checked that an "active vertex model", where active motile forces are added to the vertex model vertices, also exhibits a robust glass transition characterized by the shape order parameter $q$. The fact that two models with ostensibly different degrees of freedom share the same transition suggests that there is a deeper universality, perhaps generated by isostaticity, that remains to be understood.

Another result of this work is the surprising and unexpected differences between confluent models (such as the vertex and SPV models) and particle-based models (such as Lennard-Jones glasses and SPP models).  For example, works by Berthier~\cite{Berthier_PRL_2014}, Fily and Marchetti~\cite{Fily_2012} in SPP models suggest that the location of the zero motility glass transition packing density $\phi_G$ (defined as the density at which dynamics cease in the limit of $v_0\to0$) depends the value of noise, $D_r$. This is also related to the observation that the jamming  and glass transition are not controlled by the same critical point in non-active systems~\cite{Ikeda_PRL_2012,Teitel_athermal_vs_glass}.  We find this is not the case in the SPV model. Fig.~\ref{fig:dr}(a \& b) show that while the glass transition point $p_0^*$ shifts with $D_r$ at finite values of $v_0$, in the limit of vanishing motility, all glass transition lines merge on to a single point in the limit $v_0 \to 0$, namely $p_0^*=3.81$.

Given these differences, it is important to ask which type of model is appropriate for a given system. We  argue that SPV models are maybe more appropriate for many biological tissues. Whereas SPP models interact with two-body interactions that only depend on particle center positions, both SPV and vertex models  naturally incorporate contractility as a key property of living cells and capture the inherently multi-body nature of intercellular forces due to shape deformations.  Unlike equilibrium vertex models, SPV models account for cell motility, and they are also much easier to simulate in 3D (which is nearly impossible in practice for the vertex model.)

Recent work by Li and Sun~\cite{Li_Sun_Biophys} also  models a confluent  cell as a Voronoi tessellation of the plane. An important difference between this work and ours is that in Ref.~~\cite{Li_Sun_Biophys} 
cell-cell adhesion is captured via a potential that is quadratic in the distance between cell centers, just as in particle models. We might guess that stronger cell-cell adhesion in their model will result in stiffening of the tissue,  which is common for particle based models, although that remains to be tested in active systems. In contrast,  adhesion enters our model through the coupling of the shape energy to the cell perimeter. Increasing cell-cell adhesion (or decreasing cortical tension) yields a larger value of $p_0$, which leads to the tissue becoming softer.

 We expect that other shape-based models of confluent tissue dynamics will also yield the glass transition described here. For example, it has been reported in recent works based on the Cellular Potts model~\cite{Szabo_Potts_2010,Kabla_CPM} 
and in a modified SPP model~\cite{Garcia_Gov_2015}
 when the cell motile force is  decreased beyond a certain threshold, the motion of cells transitions from diffusive to sub-diffusive. This is similar to crossing the glass transition line in the SPV model by decreasing the value of $v_0$.

In this work and in previous work based on the vertex model, the cell volume is generally assumed to be fixed. While this is a good assumption in developmental systems such as drosopholia~\cite{Zallen_Karen,Gelbart_PNAS_2012} and zebrafish~\cite{manning_2010},   epithelial tissues cells can show  significant volume fluctuations, as reported recently ~\cite{Zehnder_Angelini_Biophys_2015,Zehnder_Angelini_PRE_2015}. Therefore, it will be important to incorporate volume fluctuations in future iterations of  the vertex model or the SPV model, as they introduce another source of active shape fluctuation and could therefore lead to  jamming or unjamming of the tissue locally and potentially shift the location of the rigidity and glass transitions.

In our version of the SPV model, we have assumed that cell polarity is controlled by simple rotational white noise. It is also possible to include more complex mechanisms. For example, external chemical or mechanical cues could be modeled by coupling $v_0$ and $\hat{\bm n}_i$ to chemoattractant or mechanical gradients, allowing waves or other pattern formation mechanisms to interact with the jamming transition. Similarly, simple alignment rules (such as those in the Viscek model~\cite{Vicsek_1995}) could lead to collective flocking modes that also affect glassy dynamics. 

Another interesting extension of the SPV model would be to study the role of cell-cell friction,
 which has already been shown to be important in controlling collective dynamics in particle-based tissue models~\cite{ Garcia_Gov_2015}.
 Our current model includes viscous frictional coupling of cell to the 2D substrate and cell-cell adhesion enters as a negative line tension on interfaces. However, it would be possible to add a frictional force between cells proportional to the length of the edge shared between two cells, and we know from previous work on particulate glasses that these localized frictions can change the  location of jamming/glass transition  and the nature of spatial correlations in a glass~\cite{Silbert_frictional_jamming, Henkes_frictional_jamming}.

It is also tempting to speculate about the relationship between the unjamming transition captured by our model and the epithelial-mesenchimal transition (EMT) that precedes cell escape from a solid tumor mass. The EMT involves significant changes in cell-cell adhesion and cytoskeletal composition, with associated changes in cell shape and motility. This suggests that escape from the tumor mass is controlled not just by the chemical breakdown of the basement membrane, but also by specific changes in mechanical properties of both individual cells and the surrounding tissue~\cite{Wong_Nat_Mat_2014}. One could then hypothesize that the collective unjamming described here may provide the first necessary step towards the mechanical changes needed for cell escape from primary tumors.

In particular, recent work suggests that cancer tumors are mechanically heterogeneous, with mixtures of stiff and soft cells that have varying degrees of active contractility~\cite{Wetzel_draft_2015}. Our jamming phase diagram suggests that the soft cells, which often exhibit mesenchymal markers and presumably correspond to higher values of $p_0$, might unjam and move towards the boundary of a primary tumor more easily than their stiff counterparts. Examining the effects of tissue heterogeneity on tissue rigidity and patterns of cell motility is therefore a very promising avenue for developing predictive theories for tumor invasiveness and metastasis.

\appendix
\section{Simulation algorithm for the SPV model}
\label{spv_forces}

To create an initial configuration for the simulation, we first generate a seed point pattern using random sequential addition (RSA)~\cite{RSA_Torquato} and anneal it by integrating Eq.~\ref{equation_of_motion} with $v_0=0$ for $100$ MD steps. The resulting structure then serves as an initially state for all simulations runs. The use of (RSA) only serves to speed up the initial seed generation as using a Poisson random point pattern does not change the results presented in this paper. 

At each time step of the simulation, a Voronoi tesselation is created based on the cell centers. The intercellular forces are then calculated based on shapes and topologies of the Voronoi cells (see discussion below). 
We employ Euler's method to carry out the numerical integration of Eq.~\ref{equation_of_motion}, i.e., at each time step of the simulation the intercellular forces is calculated  based on the cell center positions in the previous time step.

\begin{figure}[h]
\begin{center}
\includegraphics[width=0.5\columnwidth]{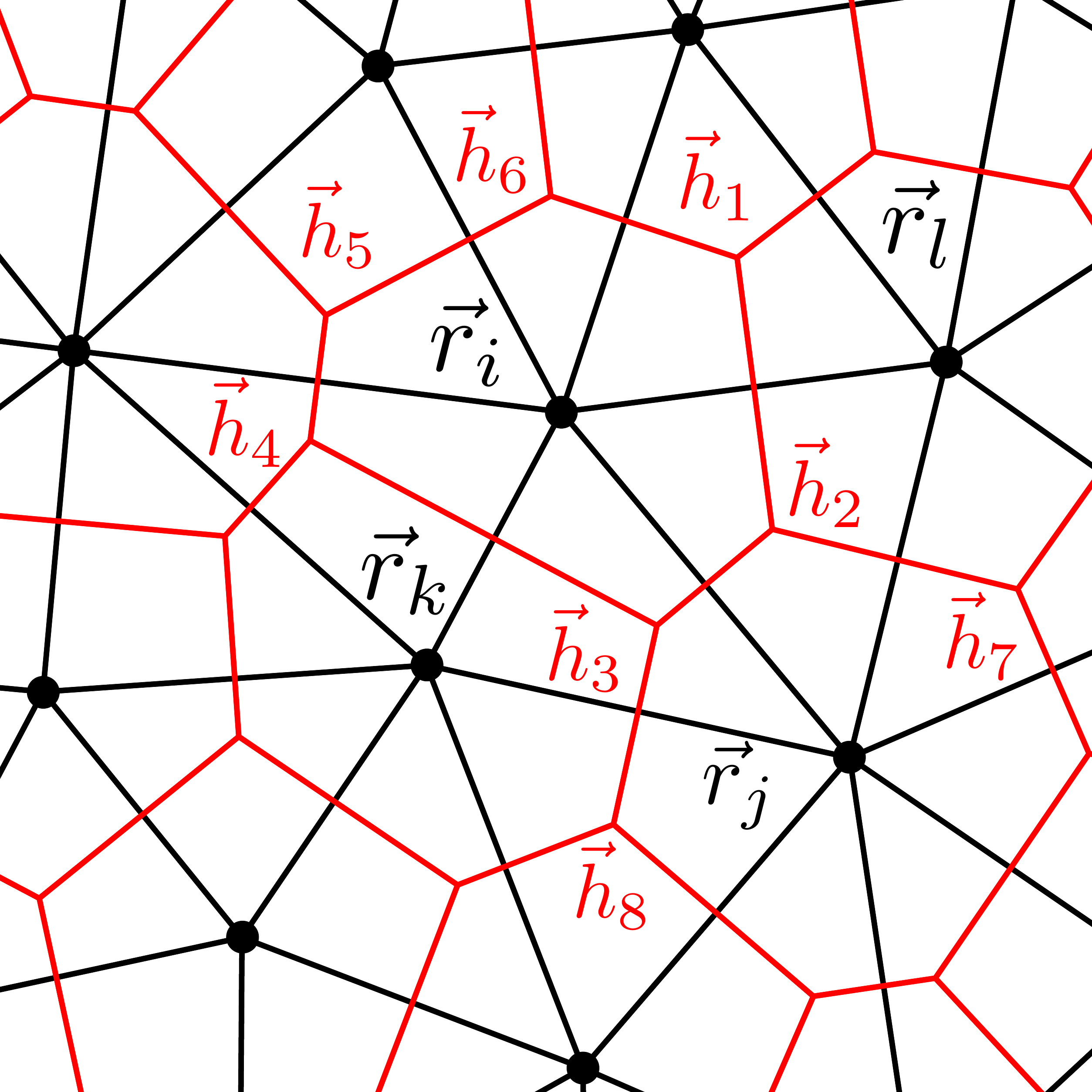}
\caption{ 
Cell centers positions are specified by vectors $\{\vec{r}\}$. They form a Delaunay triangulation (black lines). Its dual is the Voronoi tessellation (red lines), with vertices given by $\{\vec{h}\}$.
}
\label{fig:voronoi}
\end{center}
\end{figure}

In a Delaunay triangulation, a trio of neighboring Voronoi centers define a vertex of a Voronoi polygon. 
For example in Fig.~\ref{fig:voronoi}, ($\vec{r}_i,  \vec{r}_j, \vec{r}_k$) define the vertex $\vec{h}_3$, which is given by
\be
\vec{h}_3 = \alpha \vec{r}_i+\beta \vec{r}_j + \gamma \vec{r}_k,
\label{eq:si:vertex_definition}
\ee
where the coefficients are given by 
\be
\begin{split}
\alpha &= {\| \vec{r}_j - \vec{r}_k  \|}^2 (\vec{r}_i - \vec{r}_j) \cdot (\vec{r}_i - \vec{r}_k) / D\\
\beta &= {\| \vec{r}_i - \vec{r}_k  \|}^2 (\vec{r}_j - \vec{r}_i) \cdot (\vec{r}_j - \vec{r}_k) / D\\
\gamma &= {\| \vec{r}_i - \vec{r}_j  \|}^2 (\vec{r}_k - \vec{r}_i) \cdot (\vec{r}_k - \vec{r}_j) / D\\
D &= 2 {\| (\vec{r}_i - \vec{r}_j)  \times (\vec{r}_j - \vec{r}_k) \|}^2.
\end{split}
\ee

In the vertex model, the total mechanical energy of a tissue depends only on the areas and perimeters of cells:
\be
E= \sum_{i=1}^N \left[ K_P (A_i-A_0)^2+ K_P (P_i-P_0)^2\right].
\label{eq:si:total_energy}
\ee
In a Voronoi tessellation, the area and perimeter of a cell $i$ can be calculated in terms of the vertex positions
\be
\begin{split}
P_i &= \sum_{m=0}^{z_i-1}   \| \vec{h}_m - \vec{h}_{m+1} \|; \\
A_i &=\frac{1}{2} \sum_{m=0}^{z_i-1}  \| \vec{h}_m \times \vec{h}_{m+1} \|,
\label{eq:si:ap}
\end{split}
\ee
where $z_i$ is the number of vertices for cell $i$ (also number of neighboring cells) and $m$ indexes the vertices. We use the convention $\vec{h}_{z_i}=\vec{h}_0$.

With these definitions, the total force on cell-$i$ can be calculated using Eq.~\ref{eq:si:total_energy}
\be
F_{i \mu} \equiv-\frac{\partial{E}}{\partial r_{i \mu}} =- \sum_{j \in n.n.(i)}\frac{\partial{E_j}}{\partial r_{i \mu}} - \frac{\partial{E_i}}{\partial r_{i \mu}},
\label{eq:si:force_ij}
\ee
here $\mu$ denotes the cartesian coordinates ($x,y$). The first term on the r.h.s. of Eq.~\ref{eq:si:force_ij} sums over all nearest neighbors of cell  $i$.  It is the force on cell $i$ due to changes in neighboring cell shapes.   The second term is the force on cell $i$ due to shape changes brought on by its own motion.

It maybe tempting to treat $\frac{\partial{E_j}}{\partial r_{i \mu}}$ as the force between cells-$i$ and $j$, but 
\be
\frac{\partial{E_j}}{\partial r_{i \mu}} \neq \frac{\partial{E_i}}{\partial r_{j \mu}}
\label{eq:si:force_caution}
\ee
since the interaction is inherently multi-cellular in nature and interactions between $i$ and $j$ also depend on $k$ and $l$ (see Fig.~\ref{fig:voronoi}). 

For the typical configuration shown in Fig.~\ref{fig:voronoi}, the first term in Eq.~\ref{eq:si:force_ij} can be expanded using the chain rule and calculated using Eq.~\ref{eq:si:vertex_definition}
\be
\begin{split}
\frac{\partial{E_j}}{\partial r_{i \mu}} 
&= 
\sum_\nu \left( \frac{\partial{E_j}}{\partial h_{2 \nu}}\frac{\partial h_{2 \nu}}{\partial r_{i \mu}} + \frac{\partial{E_j}}{\partial h_{3 \nu}}\frac{\partial h_{3 \nu}}{ \partial r_{i \mu}} \right).
\end{split}
\label{eq:si:force_chain_rule}
\ee
In Eq.~\ref{eq:si:force_chain_rule}, only terms involving $\vec{h}_2$ and $\vec{h}_3$ are kept since $E_j$ does not  depend on other vertices of cell $i$. $\nu$ is a cartesian coordinate index. The energy derivative in Eq.~\ref{eq:si:force_chain_rule} can be calculated in a straightforward way, by using Eqs.~\ref{eq:si:total_energy} and \ref{eq:si:ap} 
\be
\begin{split}
\frac{\partial{E_j}}{\partial h_{2 x}} 
=&2  K_A (A_j-A_0)\frac{\partial{A_j}}{\partial h_{2 x}} + 2K_P(P_j-P_0)\frac{\partial{P_j}}{\partial h_{2 x}}\\
=& K_A (A_j-A_0)(h_{3y}-h_{7y})  \\
& + 2K_P(P_j-P_0)
\left(
\frac{h_{2x}-h_{7x}}{\| \vec{h}_7 - \vec{h}_{2} \|}+
\frac{h_{2x}-h_{3x}}{\| \vec{h}_2 - \vec{h}_{3} \|}
\right)
 \\
\end{split}
\ee
and
\be
\begin{split}
\frac{\partial{E_j}}{\partial h_{2 y}} 
=&2 K_A(A_j-A_0)\frac{\partial{A_j}}{\partial h_{2 y}} +2K_P(P_j-P_0)\frac{\partial{P_j}}{\partial h_{2 y}}\\
=& K_A(A_j-A_0)(h_{3x}-h_{7x})  \\
& + 2K_P(P_j-P_0)
\left(
\frac{h_{2y}-h_{7y}}{\| \vec{h}_7 - \vec{h}_{2} \|}+
\frac{h_{2y}-h_{3y}}{\| \vec{h}_2 - \vec{h}_{3} \|}
\right).
 \\
\end{split}
\ee
Similarly, the second term on the r.h.s. of Eq.~\ref{eq:si:force_ij} can be calculated in a similar way.

\section{Cell displacements and structural order parameter as a function of $D_r$}
\subsection{Expanding cell displacements in an eigenbasis associated with the underlying dynamical matrix}
\label{soft_mode_analysis}
In the absence of activity ($v_0=0$), the tissue is a solid for $p_0<p_0^*=3.81$. As $v_0$ is increased, the solid behavior persists up to $v_0=v_0^*(p_0)$, which is given by the glass transition line in Fig.~2. In order for the tissue to flow, sufficient energy input is needed to overcome energy barriers in the potential energy landscape, which are a property of the underlaying  solid state at $v_0=0$. 
In this limit, the instantaneous cell center positions $\{ \vec{r}_{i}(t) \}$  can be thought of as a small displacement $\{ \vec{d}_i(t)  \}$ from the nearest solid reference state $\{ \vec{r}_{0i} \}$~\cite{Henkes2011} where $\vec{d}_i(t) =\vec{r}_{i} - \vec{r}_{0i}$. 
The $\vec{r}_{0i}$ correspond to positions of cell in a solid, which has a well-defined linear response regime~\cite{bi_manning_nphys}. The linear response is most conveniently expressed as the eigen-spectrum of the dynamical matrix $D_{ij\alpha\beta}$. Since the eigenvectors $\{ \hat{e}_{i,\nu}\}$ of $D_{ij\alpha\beta}$ form a complete orthonormal basis, the cell center displacement can then be expressed as a linear combination of $\{ \hat{e}_{i,\nu}\}$
\be
\vec{d}_i(t) = \sum_\nu a_\nu(t)  \hat{e}_{i,\nu}
\label{dvec_expand}
\ee
For simplicity, we will adopt the Bra-ket notation and express the eigenbasis simply as $\ket{\nu}$ and  Eq.~\ref{dvec_expand} becomes
\be
\ket{d} =\sum_\nu a_\nu(t) \ket{\nu},
\label{ket_d}
\ee
where 
\be
\hat{D} \ket{\nu} = \omega_{\nu}^2 \ket{\nu} 
\label{eigenbasis}
\ee
and $\omega_{\nu}^2$ are the eigenvalues of the dynamical matrix. 

The polarization  vector $\hat{n}_i$ can also be expressed as a linear combination of eigenvectors
\be
\ket{n} = \sum_\nu b_\nu(t) \ket{\nu}.
\label{ket_n}
\ee
Since the polarization vector and eigenvector are both unit vectors, 
it follows that $b_\nu(t)=\braket{n}{\nu}=cos(\theta_\nu-\psi) $. Where $\psi$ is the angle of the polarization and $\theta_\nu$ the angle of the eigenvector. 

Then the equation of motion for $\vec{d}_i$ (Eq.~2), can be rewritten as
\be
\dot{\vec{d}} = -\mu\frac{\partial E}{\partial \vec{r}_i } \bigg|_{\vec{r}_{0,i}} +v_0 \hat{n}_i
\label{eom_d}
\ee
Using Eqs.~\ref{ket_d}--\ref{eom_d}, we find
\be
\begin{split}
 \frac{d}{dt}\braket{\nu}{d} &= - \mu \braket{\nu}{\hat{D} \ d} +v_0 \braket{\nu}{n} \ \ ,\text{or} \\
 \frac{d}{dt}a_\nu(t) &= - \mu \omega_\nu^2 a_\nu(t)+v_0 b_\nu(t).
\end{split}
\label{eom_ket}
\ee

Then the equation of motion for each amplitude is
\be
\begin{split}
\frac{d}{dt}a_\nu(t) &= - \mu \omega_\nu^2 a_\nu(t)+v_0 cos(\theta_\nu-\psi) \\
\dot{\psi} &= \eta.
\end{split}
\ee
This is just the equation of motion for a self-propelled particle tethered to a spring with active forcing that is strongest along the direction of the eigenvector~\cite{Gov_elastic_gel_PRE_2015}. The solution is then:

\be
a_\nu(t) =a_\nu(t=0)e^{-kt} +  v_0 \int_{0}^{t} dt' e^{-k(t-t')} cos(\theta_\nu-\psi),
\ee
where $k=\mu \omega_\nu^2$.

Solving for the ensemble averaged quantity:
\be
\langle a_\nu(t) \rangle =a_\nu(t=0)e^{-kt} + v_0 \int_{0}^{t} dt' e^{-k(t-t')} 
\langle cos(\theta_\nu-\psi) \rangle,
\ee
 and using the relations
\be
\begin{split}
\langle \cos \psi(t)  \rangle &= \cos \psi(0) e^{-D_r t}; \\
\langle \sin \psi(t)  \rangle &= \sin \psi(0) e^{-D_r t} \\
\cos(\theta_\nu-\psi) &=\sin(\theta_\nu)\sin(\psi)+\cos(\theta_\nu)\cos(\psi),
\end{split}
\ee
to get the ensemble averaged solution for the amplitude becomes
\be
\langle a_\nu(t) \rangle =a_\nu(0)e^{-kt} + v_0 \cos\left(\theta_\nu-\psi(0)\right)
\frac{e^{-kt} - e^{-D_r t}}{D_r -k}.
\label{amplitude_solution}
\ee
In the  limit of $D_r \to \infty$ and Eq.~\ref{amplitude_solution} becomes 
\be
\langle a_\nu(t) \rangle =a_\nu(0)e^{-kt} .
\ee
This suggests that while normal modes control the rate of decay, they do no affect the long-time behavior.  
 
However as $D_r \to 0$, Eq.~\ref{amplitude_solution} becomes
\be
a_\nu(t) = a_\nu(0)e^{-\mu \omega_\nu^2t} +\frac{v_0}{\mu \omega_\nu^2} \cos\left(\theta_\nu-\psi(0)\right)  \left(1-e^{-\mu \omega_\nu^2 t}\right)
\ee
The second term in this equation scales as $\sim 1/\omega_\nu^2$. Therefore, at short times (corresponding to instantaneous response), the mode amplitude $a_\nu$
is much larger for modes at lower frequencies. 
Since the reference state is an elastic solid with Debye scaling $D(\omega) \sim \omega$ as $\omega \to 0$~\cite{bi_manning_nphys}, this suggests that the displacement will be  heavily dominated by the  lowest frequency modes that are spatially more collective in nature. 

\subsection{Effect of $D_r$ on glass transition boundary}
\label{Dr_dep_boundary}
\begin{figure}[h]
\begin{center}
\includegraphics[width=1\columnwidth]{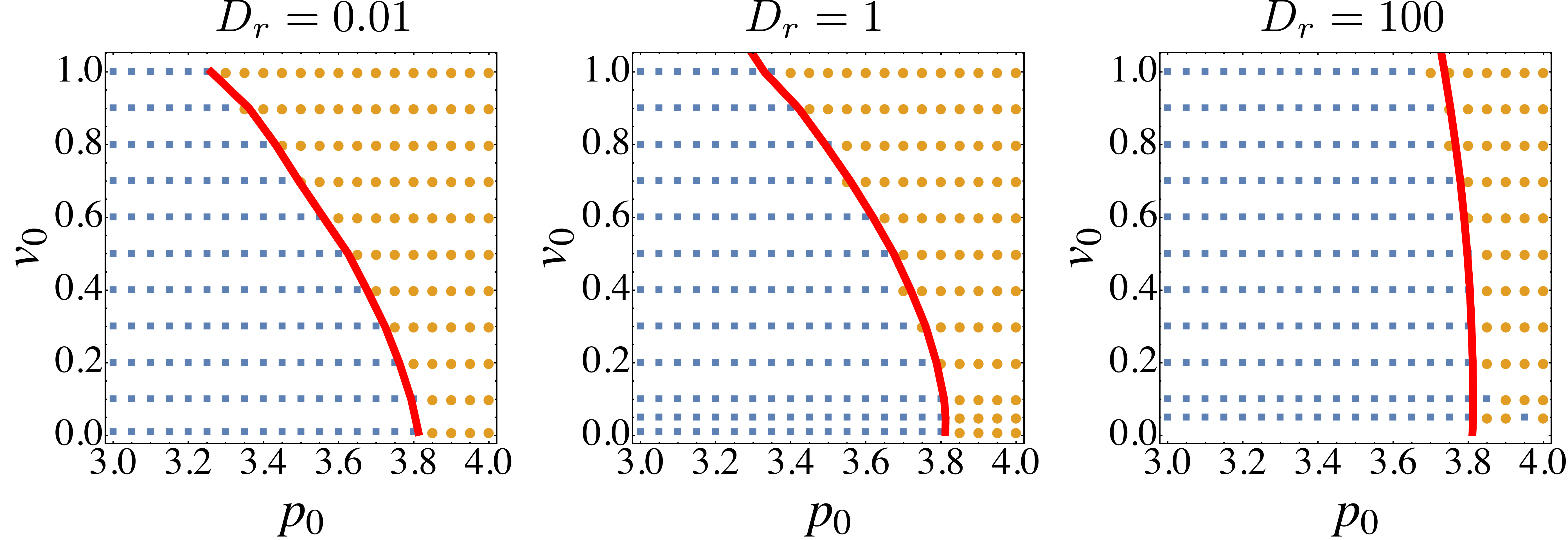}
\caption{ 
Comparison between glass transition boundaries obtained using shape order parameter (red line) and $D_{eff}$ (blue squares and orange circles). }
\label{fig:boundary_comparison}
\end{center}
\end{figure}

Figure.~\ref{fig:boundary_comparison} shows that the location in phase space where the shape index $q=3.81$ is in excellent agreement with the dynamical solid-fluid phase boundary determined by $D_{eff}$, at all values of $D_r$.  

\begin{acknowledgments}
M.L.M. acknowledges support from the Alfred P. Sloan Foundation. 
M.L.M and D.B. acknowledge support from NSF-BMMB-1334611 and NSF-DMR-1352184,
	the Gordon and Betty Moore Foundation and
	the Research Corporation for Scientific Advancement and 
	NIH R01GM117598-02. 
M.C.M. acknowledges support from the Simons Foundation. 
M.C.M. and X.Y. acknowledge support from NSF-DMR-305184. 
The authors also acknowledge the Syracuse University HTC Campus Grid, NSF award ACI-1341006 and the Soft Matter Program at Syracuse University.  
\end{acknowledgments}
\bibliographystyle{plain}

\bibliography{motility}

\begin{thebibliography}{10}

\bibitem{Abate_Durian_effective_temp}
A.~R. Abate and D.~J. Durian.
\newblock Effective temperatures and activated dynamics for a two-dimensional
  air-driven granular system on two approaches to jamming.
\newblock {\em Phys. Rev. Lett.}, 101:245701, Dec 2008.

\bibitem{Angelini_PRL_2010}
Thomas~E. Angelini, Edouard Hannezo, Xavier Trepat, Jeffrey~J. Fredberg, and
  David~A. Weitz.
\newblock Cell migration driven by cooperative substrate deformation patterns.
\newblock {\em Phys. Rev. Lett.}, 104:168104, Apr 2010.

\bibitem{Angelini_PNAS_2011}
Thomas~E. Angelini, Edouard Hannezo, Xavier Trepat, Manuel Marquez, Jeffrey~J.
  Fredberg, and David~A. Weitz.
\newblock Glass-like dynamics of collective cell migration.
\newblock {\em Proceedings of the National Academy of Sciences},
  108(12):4714--4719, 2011.

\bibitem{Belmonte_PRL_2008_cell_sorting}
Julio~M. Belmonte, Gilberto~L. Thomas, Leonardo~G. Brunnet, Rita M.~C.
  de~Almeida, and Hugues Chat\'e.
\newblock Self-propelled particle model for cell-sorting phenomena.
\newblock {\em Phys. Rev. Lett.}, 100:248702, Jun 2008.

\bibitem{Gov_elastic_gel_PRE_2015}
E.~Ben-Isaac, \'E. Fodor, P.~Visco, F.~van Wijland, and Nir~S. Gov.
\newblock Modeling the dynamics of a tracer particle in an elastic active gel.
\newblock {\em Phys. Rev. E}, 92:012716, Jul 2015.

\bibitem{Berthier_PRL_2014}
Ludovic Berthier.
\newblock Nonequilibrium glassy dynamics of self-propelled hard disks.
\newblock {\em Physical Review Letters}, 112(22):220602, 2014.

\bibitem{bi_manning_nphys}
Dapeng Bi, J.~H. Lopez, J.~M. Schwarz, and M.~Lisa Manning.
\newblock A density-independent rigidity transition in biological tissues.
\newblock {\em Nat Phys}, advance online publication:--, 09 2015.

\bibitem{bi_softmatter}
Dapeng Bi, Jorge~H. Lopez, J.~M. Schwarz, and M.~Lisa Manning.
\newblock Energy barriers and cell migration in densely packed tissues.
\newblock {\em Soft Matter}, 10:1885--1890, 2014.

\bibitem{Cecconi_PRL_2003}
Fabio Cecconi, Andrea Puglisi, Umberto Marini~Bettolo Marconi, and Angelo
  Vulpiani.
\newblock Noise activated granular dynamics.
\newblock {\em Phys. Rev. Lett.}, 90:064301, Feb 2003.

\bibitem{Deforet_ncomm_2014}
M.~Deforet, V.~Hakim, H.~G. Yevick, G.~Duclos, and P.~Silberzan.
\newblock Emergence of collective modes and tri-dimensional structures from
  epithelial confinement.
\newblock {\em Nat Commun}, 5, 05 2014.

\bibitem{matthias_elife}
Raphaël Etournay, Marko Popović, Matthias Merkel, Amitabha Nandi, Corinna
  Blasse, Benoît Aigouy, Holger Brandl, Gene Myers, Guillaume Salbreux, Frank
  Jülicher, and Suzanne Eaton.
\newblock Interplay of cell dynamics and epithelial tension during
  morphogenesis of the \textit{Drosophila} pupal wing.
\newblock {\em eLife}, 4:e07090, jun 2015.

\bibitem{Farhadifar2007}
Reza Farhadifar, Jens-Christian Roeper, Benoit Aigouy, Suzanne Eaton, and Frank
  Julicher.
\newblock The influence of cell mechanics, cell-cell interactions, and
  proliferation on epithelial packing.
\newblock {\em Current Biology}, 17(24):2095 -- 2104, 2007.

\bibitem{Fily_SM_2014}
Yaouen Fily, Silke Henkes, and M.~Cristina Marchetti.
\newblock Freezing and phase separation of self-propelled disks.
\newblock {\em Soft Matter}, 10:2132--2140, 2014.

\bibitem{Fily_2012}
Yaouen Fily and M.~Cristina Marchetti.
\newblock Athermal phase separation of self-propelled particles with no
  alignment.
\newblock {\em Phys. Rev. Lett.}, 108:235702, Jun 2012.

\bibitem{Garcia_Gov_2015}
Simon Garcia, Edouard Hannezo, Jens Elgeti, Jean-François Joanny, Pascal
  Silberzan, and Nir~S. Gov.
\newblock Physics of active jamming during collective cellular motion in a
  monolayer.
\newblock {\em Proceedings of the National Academy of Sciences},
  112(50):15314--15319, 2015.

\bibitem{Gelbart_PNAS_2012}
Michael~A. Gelbart, Bing He, Adam~C. Martin, Stephan~Y. Thiberge, Eric~F.
  Wieschaus, and Matthias Kaschube.
\newblock Volume conservation principle involved in cell lengthening and
  nucleus movement during tissue morphogenesis.
\newblock {\em Proceedings of the National Academy of Sciences},
  109(47):19298--19303, 2012.

\bibitem{MET_review}
N.P.A.Devika Gunasinghe, Alan Wells, ErikW. Thompson, and HonorJ. Hugo.
\newblock Mesenchymal–epithelial transition (met) as a mechanism for
  metastatic colonisation in breast cancer.
\newblock {\em Cancer and Metastasis Reviews}, 31(3-4):469--478, 2012.

\bibitem{Friedl_2014}
Anna Haeger, Marina Krause, Katarina Wolf, and Peter Friedl.
\newblock Cell jamming: Collective invasion of mesenchymal tumor cells imposed
  by tissue confinement.
\newblock {\em Biochimica et Biophysica Acta (BBA) - General Subjects},
  1840(8):2386 -- 2395, 2014.
\newblock Matrix-mediated cell behaviour and properties.

\bibitem{Henkes_frictional_jamming}
S.~Henkes, M.~van Hecke, and W.~van Saarloos.
\newblock Critical jamming of frictional grains in the generalized isostaticity
  picture.
\newblock {\em EPL (Europhysics Letters)}, 90(1):14003, 2010.

\bibitem{Henkes2011}
Silke Henkes, Yaouen Fily, and M.~Cristina Marchetti.
\newblock Active jamming: Self-propelled soft particles at high density.
\newblock {\em Phys. Rev. E}, 84:040301, Oct 2011.

\bibitem{honda_voronoi}
Hisao Honda.
\newblock Description of cellular patterns by dirichlet domains: The
  two-dimensional case.
\newblock {\em Journal of Theoretical Biology}, 72(3):523 -- 543, 1978.

\bibitem{Hufnagel2007}
Lars Hufnagel, Aurelio~A. Teleman, H.~Rouault, Stephen~M. Cohen, and Boris~I.
  Shraiman.
\newblock On the mechanism of wing size determination in fly development.
\newblock {\em Proceedings of the National Academy of Sciences},
  104(10):3835--3840, March 2007.

\bibitem{Ikeda_PRL_2012}
Atsushi Ikeda, Ludovic Berthier, and Peter Sollich.
\newblock Unified study of glass and jamming rheology in soft particle systems.
\newblock {\em Phys. Rev. Lett.}, 109:018301, Jul 2012.

\bibitem{Kabla_CPM}
Alexandre~J. Kabla.
\newblock Collective cell migration: leadership, invasion and segregation.
\newblock {\em Journal of The Royal Society Interface}, 9(77):3268--3278, 2012.

\bibitem{Zallen_Karen}
Karen~E Kasza, Dene~L Farrell, and Jennifer~A Zallen.
\newblock Spatiotemporal control of epithelial remodeling by regulated myosin
  phosphorylation.
\newblock {\em Proc Natl Acad Sci U S A}, 111(32):11732--11737, Aug 2014.

\bibitem{Kim_monoloayer_obstacle_2013}
Jae~Hun Kim, Xavier Serra-Picamal, Dhananjay~T. Tambe, Enhua~H. Zhou,
  Chan~Young Park, Monirosadat Sadati, Jin-Ah Park, Ramaswamy Krishnan, Bomi
  Gweon, Emil Millet, James~P. Butler, Xavier Trepat, and Jeffrey~J. Fredberg.
\newblock Propulsion and navigation within the advancing monolayer sheet.
\newblock {\em Nat Mater}, 12(9):856--863, 09 2013.

\bibitem{Dirichlet_1850}
G.~Lejeune~Dirichlet.
\newblock Über die reduction der positiven quadratischen formen mit drei
  unbestimmten ganzen zahlen.
\newblock {\em Journal fur die reine und angewandte Mathematik}, 40:209--227,
  1850.

\bibitem{Li_Sun_Biophys}
Bo~Li and Sean~X. Sun.
\newblock Coherent motions in confluent cell monolayer sheets.
\newblock {\em Biophysical Journal}, 107(7):1532--1541, 2015/08/23 2015.

\bibitem{LiuNagelReview}
Andrea~J. Liu and Sidney~R. Nagel.
\newblock The jamming transition and the marginally jammed solid.
\newblock {\em Annual Review of Condensed Matter Physics}, 1(1):347--369, 2010.

\bibitem{Manning_Liu_PRL_2011}
M.~L. Manning and A.~J. Liu.
\newblock Vibrational modes identify soft spots in a sheared disordered
  packing.
\newblock {\em Phys. Rev. Lett.}, 107:108302, Aug 2011.

\bibitem{manning_2010}
M.~Lisa Manning, Ramsey~A. Foty, Malcolm~S. Steinberg, and Eva-Maria Schoetz.
\newblock Coaction of intercellular adhesion and cortical tension specifies
  tissue surface tension.
\newblock {\em Proceedings of the National Academy of Sciences},
  107(28):12517--12522, 07 2010.

\bibitem{bouchaud_trap}
Cecile Monthus and Jean-Philippe Bouchaud.
\newblock Models of traps and glass phenomenology.
\newblock {\em Journal of Physics A: Mathematical and General}, 29(14):3847,
  1996.

\bibitem{ngai_honda}
Tatsuzo Nagai and Hisao Honda.
\newblock A dynamic cell model for the formation of epithelial tissues.
\newblock {\em Philosophical Magazine Part B}, 81(7):699--719, 2001.

\bibitem{Nakaya_MET}
Yukiko Nakaya, Shinya Kuroda, Yuji~T. Katagiri, Kozo Kaibuchi, and Yoshiko
  Takahashi.
\newblock Mesenchymal-epithelial transition during somitic segmentation is
  regulated by differential roles of cdc42 and rac1.
\newblock {\em Developmental Cell}, 7(3):425--438, 2015.

\bibitem{Ran_2013}
Ran Ni, Martien A.~Cohen Stuart, and Marjolein Dijkstra.
\newblock Pushing the glass transition towards random close packing using
  self-propelled hard spheres.
\newblock {\em Nat Commun}, 4, 10 2013.

\bibitem{KaesNJP}
Kenechukwu~David Nnetu, Melanie Knorr, Josef K{\"a}s, and Mareike Zink.
\newblock The impact of jamming on boundaries of collectively moving
  weak-interacting cells.
\newblock {\em New Journal of Physics}, 14(11):115012, 2012.

\bibitem{Teitel_athermal_vs_glass}
Peter Olsson and S.~Teitel.
\newblock Athermal jamming versus thermalized glassiness in sheared
  frictionless particles.
\newblock {\em Phys. Rev. E}, 88:010301, Jul 2013.

\bibitem{Park_2015}
Jin-Ah Park, Jae~Hun Kim, Dapeng Bi, Jennifer~A. Mitchel, Nader~Taheri Qazvini,
  Kelan Tantisira, Chan~Young Park, Maureen McGill, Sae-Hoon Kim, Bomi Gweon,
  Jacob Notbohm, Robert Steward~Jr, Stephanie Burger, Scott~H. Randell,
  Alvin~T. Kho, Dhananjay~T. Tambe, Corey Hardin, Stephanie~A. Shore, Elliot
  Israel, David~A. Weitz, Daniel~J. Tschumperlin, Elizabeth~P. Henske, Scott~T.
  Weiss, M.~Lisa Manning, James~P. Butler, Jeffrey~M. Drazen, and Jeffrey~J.
  Fredberg.
\newblock Unjamming and cell shape in the asthmatic airway epithelium.
\newblock {\em Nat Mater}, 14(10):1040--1048, 10 2015.

\bibitem{Silberzan2010}
L~Petitjean, M~Reffay, E~Grasland-Mongrain, M~Poujade, B~Ladoux, A~Buguin, and
  P~Silberzan.
\newblock Velocity fields in a collectively migrating epithelium.
\newblock {\em Biophysical journal}, 98(9):1790--1800, 2010.

\bibitem{Poujade_2007}
M.~Poujade, E.~Grasland-Mongrain, A.~Hertzog, J.~Jouanneau, P.~Chavrier,
  B.~Ladoux, A.~Buguin, and P.~Silberzan.
\newblock Collective migration of an epithelial monolayer in response to a
  model wound.
\newblock {\em Proceedings of the National Academy of Sciences},
  104(41):15988--15993, 10 2007.

\bibitem{Puliafito_2012}
Alberto Puliafito, Lars Hufnagel, Pierre Neveu, Sebastian Streichan, Alex
  Sigal, D.~Kuchnir Fygenson, and Boris~I. Shraiman.
\newblock Collective and single cell behavior in epithelial contact inhibition.
\newblock {\em Proceedings of the National Academy of Sciences},
  109(3):739--744, 2012.

\bibitem{Sadati-Fredberg-review}
Monirosadat Sadati, Nader~Taheri Qazvini, Ramaswamy Krishnan, Chan~Young Park,
  and Jeffrey~J. Fredberg.
\newblock Collective migration and cell jamming.
\newblock {\em Differentiation}, 86(3):121 -- 125, 2013.
\newblock Mechanotransduction.

\bibitem{Schoetz2008}
E.-M. Schoetz, R.~D. Burdine, F.~Julicher, M.~S. Steinberg, C.-P. Heisenberg,
  and R.~A. Foty.
\newblock Quantitative differences in tissue surface tension influence
  zebrafish germlayer positioning.
\newblock {\em HFSP journal}, Vol.2 (1):1--56, 2008.

\bibitem{Schoetz2013}
Eva-Maria Schoetz, Marcos Lanio, Jared~A. Talbot, and M.~Lisa Manning.
\newblock Glassy dynamics in three-dimensional embryonic tissues.
\newblock {\em J. Roy. Soc. Interface}, 10:20130726, 2013.

\bibitem{Sepulvieda_et_al}
Nestor Sepulveda, Laurence Petitjean, Olivier Cochet, Erwan Grasland-Mongrain,
  Pascal Silberzan, and Vincent Hakim.
\newblock Collective cell motion in an epithelial sheet can be quantitatively
  described by a stochastic interacting particle model.
\newblock {\em PLoS Comput Biol}, 9(3):e1002944, 03 2013.

\bibitem{Silbert_frictional_jamming}
Leonardo~E. Silbert.
\newblock Jamming of frictional spheres and random loose packing.
\newblock {\em Soft Matter}, 6:2918--2924, 2010.

\bibitem{Sollich_SGR_1998}
Peter Sollich.
\newblock Rheological constitutive equation for a model of soft glassy
  materials.
\newblock {\em Phys. Rev. E}, 58:738--759, Jul 1998.

\bibitem{Gupta_etal_2015}
S~S Soumya, Animesh Gupta, Andrea Cugno, Luca Deseri, Kaushik Dayal, Dibyendu
  Das, Shamik Sen, and Mandar~M. Inamdar.
\newblock Coherent motion of monolayer sheets under confinement and its
  pathological implications.
\newblock {\em PLoS Comput Biol}, 11(12):e1004670, 12 2015.

\bibitem{Staple2010}
D.~B Staple, R~Farhadifar, J.~C Roeper, B~Aigouy, S~Eaton, and F~Julicher.
\newblock Mechanics and remodelling of cell packings in epithelia.
\newblock {\em Eur. Phys. J. E}, 33(2):117--127, Oct 17 2010.

\bibitem{Szabo_Potts_2010}
A~Szabo, Runnep, E~Mehes, W~O Twal, W~S Argraves, Y~Cao, and A~Czirok.
\newblock Collective cell motion in endothelial monolayers.
\newblock {\em Physical Biology}, 7(4):046007, 2010.

\bibitem{Szabo_PRE_2006}
B.~Szabo, G.~J. Sz\"oll\"osi, B.~G\"onci, Zs. Jur\'anyi, D.~Selmeczi, and
  Tam\'as Vicsek.
\newblock Phase transition in the collective migration of tissue cells:
  Experiment and model.
\newblock {\em Phys. Rev. E}, 74:061908, Dec 2006.

\bibitem{Theiry_EMT_review}
Jean~Paul Thiery.
\newblock Epithelial-mesenchymal transitions in tumour progression.
\newblock {\em Nat Rev Cancer}, 2(6):442--454, 06 2002.

\bibitem{Thompson_Newgreen_EMT_review}
Erik~W. Thompson and Donald~F. Newgreen.
\newblock Carcinoma invasion and metastasis: A role for epithelial-mesenchymal
  transition?
\newblock {\em Cancer Research}, 65(14):5991--5995, 2005.

\bibitem{RSA_Torquato}
S~Torquato, Author and HW~Haslach, Jr.
\newblock Random heterogeneous materials: Microstructure and macroscopic
  properties.
\newblock {\em Applied Mechanics Reviews}, 55(4):B62--B63, 07 2002.

\bibitem{Trappe_nature_2001}
V~Trappe, V~Prasad, L~Cipelletti, and P~Segre{\ldots}.
\newblock Jamming phase diagram for attractive particles.
\newblock {\em Nature}, Jan 2001.

\bibitem{van_Hove_1954}
L\'eon Van~Hove.
\newblock Correlations in space and time and born approximation scattering in
  systems of interacting particles.
\newblock {\em Phys. Rev.}, 95:249--262, Jul 1954.

\bibitem{Vicsek_1995}
Tam\'as Vicsek, Andr\'as Czir\'ok, Eshel Ben-Jacob, Inon Cohen, and Ofer
  Shochet.
\newblock Novel type of phase transition in a system of self-driven particles.
\newblock {\em Phys. Rev. Lett.}, 75:1226--1229, Aug 1995.

\bibitem{Weaire_book}
Denis~L Weaire and Stefan Hutzler.
\newblock {\em The physics of foams}.
\newblock Oxford University Press, 1999.

\bibitem{Wetzel_draft_2015}
F.~Wetzel, A.~Fritsch, D.~Bi, R.~Stange, S.~Pawlizak, T.~Kie$\beta$ling, L-C.
  Horn, K.~Bendrat, M.~Oktay, M.~Zink, A.~Niendorf, J.~Condeelis,
  M.~H\"{o}ckel, M.~C. Marchetti, M.~L. Manning, and J.~K. Kas.
\newblock {\em Unpublished}, 2015.

\bibitem{Wong_Nat_Mat_2014}
Ian~Y. Wong, Sarah Javaid, Elisabeth~A. Wong, Sinem Perk, Daniel~A. Haber,
  Mehmet Toner, and Daniel Irimia.
\newblock Collective and individual migration following the
  epithelial--mesenchymal transition.
\newblock {\em Nat Mater}, 13(11):1063--1071, 11 2014.

\bibitem{Xu_EPL_2010}
N.~Xu, V.~Vitelli, A.~J. Liu, and S.~R. Nagel.
\newblock Anharmonic and quasi-localized vibrations in jammed solids—modes
  for mechanical failure.
\newblock {\em EPL (Europhysics Letters)}, 90(5):56001, 2010.

\bibitem{SPV_stress_paper}
X.~B. Yang, D.~Bi, M.~C. Marchetti, and M.~L. Manning.
\newblock {\em Unpublished}, 2015.

\bibitem{Zehnder_Angelini_Biophys_2015}
Steven~M. Zehnder, Melanie Suaris, Madisonclaire~M. Bellaire, and Thomas~E.
  Angelini.
\newblock Cell volume fluctuations in \{MDCK\} monolayers.
\newblock {\em Biophysical Journal}, 108(2):247 -- 250, 2015.

\bibitem{Zehnder_Angelini_PRE_2015}
Steven~M. Zehnder, Marina~K. Wiatt, Juan~M. Uruena, Alison~C. Dunn, W.~Gregory
  Sawyer, and Thomas~E. Angelini.
\newblock Multicellular density fluctuations in epithelial monolayers.
\newblock {\em Phys. Rev. E}, 92:032729, Sep 2015.

\end{thebibliography}

\end{document}